\documentclass{appolb}
\usepackage{epsfig}

\def\be{\begin{equation}}
\def\ee{\end{equation}}

\begin{document}

\title{Thermal hadron production in relativistic nuclear collisions}

\author{
A.~Andronic
\address{GSI Helmholtzzentrum f\"ur Schwerionenforschung,
D-64291 Darmstadt, Germany}
\and P.~Braun-Munzinger
\address{ExtreMe Matter Institute EMMI, GSI Helmholtzzentrum f\"ur Schwerionenforschung,
D-64291 Darmstadt, Germany and 
Technical University Darmstadt, D-64289 Darmstadt, Germany}
\and J.~Stachel
\address{Physikalisches Institut der Universit\"at Heidelberg,
D-69120 Heidelberg, Germany}
}

\maketitle

\begin{abstract}
  We present the status of the description of hadron production in central 
nucleus-nucleus collisions within the statistical model . 
Extending previous studies by inclusion of very high-mass
resonances (m$>$ 2 GeV), and the up-to-now neglected scalar $\sigma$ meson
leads to an improved description of the data. In particular, the
hitherto poorly reproduced energy dependence of the $K^+/\pi^+$ ratio at SPS
energies (``the horn'') is now well described through the connection to
the hadronic mass spectrum and, implicitly, Hagedorn's limiting temperature. 
\end{abstract}

\section{Introduction}

One of the major goals of ultrarelativistic nuclear collision studies is to
obtain information on the QCD phase diagram \cite{pbm_wambach}.  A promising
approach is the investigation of hadron production.  Hadron yields measured in
central heavy ion collisions from AGS up to RHIC energies can be described very
well 
\cite{agssps,satz,heppe,cley,beca1,rhic,nu,beca2,rapp,becgaz,aa05,man08,letessier05}
within a hadro-chemical equilibrium model.  In our approach
\cite{agssps,heppe,rhic,aa05} the only parameters are the chemical freeze-out
temperature $T$ and the baryo-chemical potential $\mu_b$ (and the fireball
volume $V$, in case yields rather than ratios of yields are fitted); for a
review see \cite{review}.

The main result of these investigations was that the extracted temperature
values rise rather sharply from low energies on towards $\sqrt{s_{NN}}\simeq$10 GeV
and reach afterwards constant values near $T$=160 MeV, while the baryochemical
potential  decreases smoothly as a function of energy.
The limiting temperature \cite{hagedorn85} behavior suggests a connection to 
the phase boundary and it was, indeed, argued \cite{wetterich} that the 
quark-hadron phase transition drives the equilibration dynamically, at least 
for  SPS energies and above. 
Considering also the results obtained for elementary collisions, where similar
analyses of hadron multiplicities, albeit with several additional, non-statistical 
parameters (see \cite{aa08,becattini08} and refs. therein),  yield 
also temperature values in the range of 160 MeV, alternative interpretations 
were put forward. These include conjectures  that the thermodynamical 
state is not reached by dynamical equilibration among constituents but rather 
is a generic fingerprint of hadronization \cite{stock,heinz}, or is a feature 
of the excited QCD vacuum \cite{castorina}. The results presented below 
lend further support to the interpretation that the phase boundary is reflected in
features of the hadron yields. 

While in general all hadron yields are described rather quantitatively, a
notable exception was up-to-now the energy dependence of the $K^+/\pi^+$ ratio which
exhibits a rather marked maximum, ``the horn'' \cite{gaz}, 
near $\sqrt{s_{NN}}\simeq$ 10 GeV \cite{na49pi}. 
Predicted first within a model of quark-gluon plasma (QGP) formation \cite{gaz},
the existence of such a maximum was also predicted \cite{pbm4} within 
the framework of the statistical model, but the observed rather sharp structure 
could not be reproduced \cite{aa05}. 
Other attempts to describe the energy dependence of the $K^+/\pi^+$ ratio within 
the thermal model \cite{becgaz,letessier05} were also not successful, except 
when an energy-dependent light quark fugacity was used as an additional parameter
\cite{letessier05}.
Furthermore, all attempts to reproduce this structure
within the framework of hadronic cascade models also failed, as is discussed 
in detail in \cite{na49pi}. As a consequence, the horn structure is taken in
\cite{na49pi} as experimental evidence for the onset of deconfinement and
QGP formation, and as support for the predictions of \cite{gaz}.
We have recently shown \cite{aat2} that, employing an improved hadronic mass 
spectrum, in which the $\sigma$ meson \cite{gar07} and many higher-lying resonances 
are included, leads to an increase of about 16\% for the calculated 
pion yields. This increase levels off near the point where the temperature reaches 
its limiting value, thereby sharpening the structure in the $K^+/\pi^+$ ratio, 
as will be shown below.
For the $\Lambda$ hyperons, the new high mass resonances lead to an increase 
in the calculated production of about 22\%.
An increase of up to 6\% is observed for protons, while for kaons this increase 
is about 7\%.

\begin{figure}[hbt]
\begin{tabular}{lr} \begin{minipage}{.49\textwidth}
\hspace{-.3cm}\includegraphics[width=1.\textwidth]{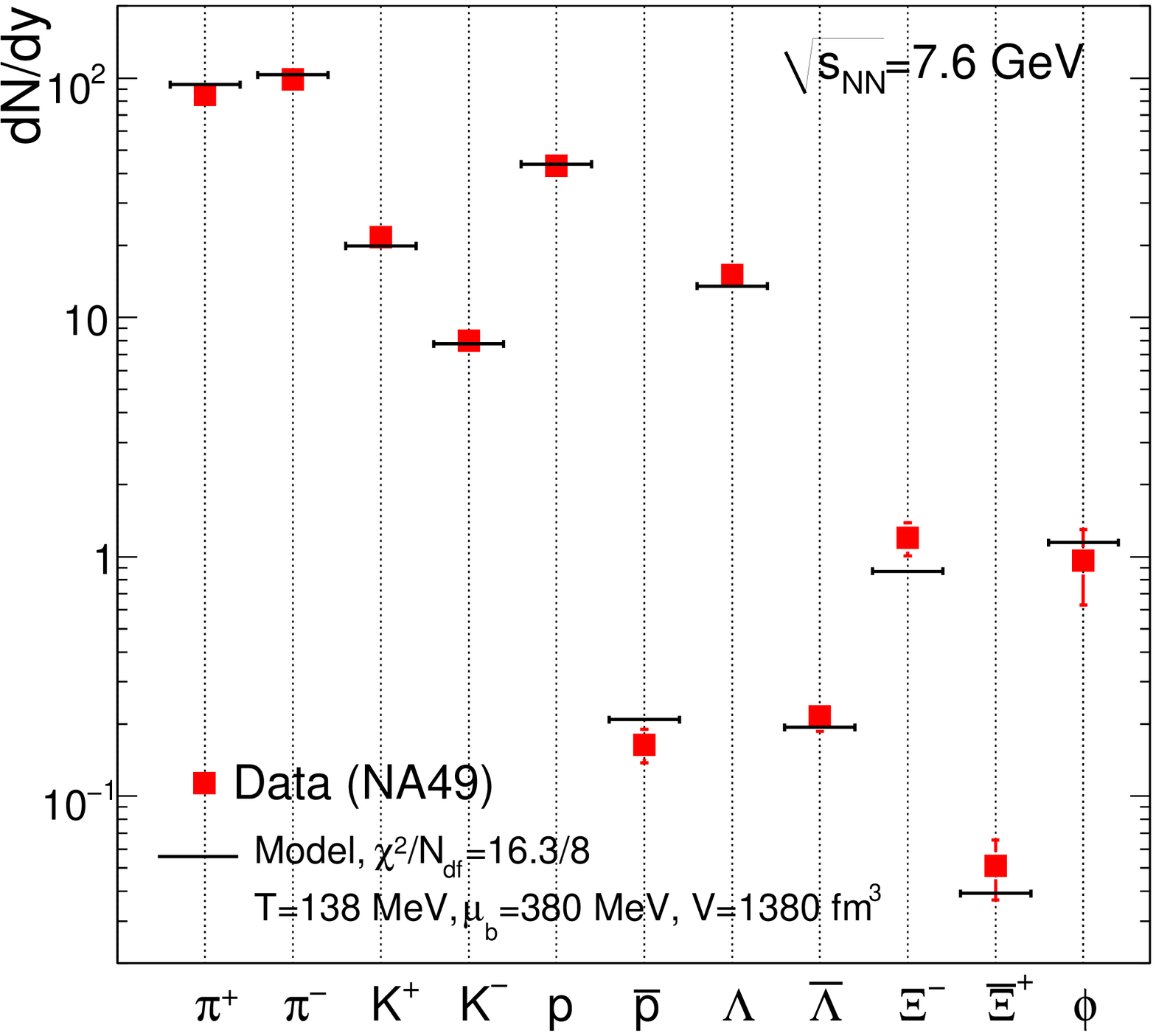}
\end{minipage} &\begin{minipage}{.49\textwidth}
\hspace{-.5cm}\includegraphics[width=1.\textwidth]{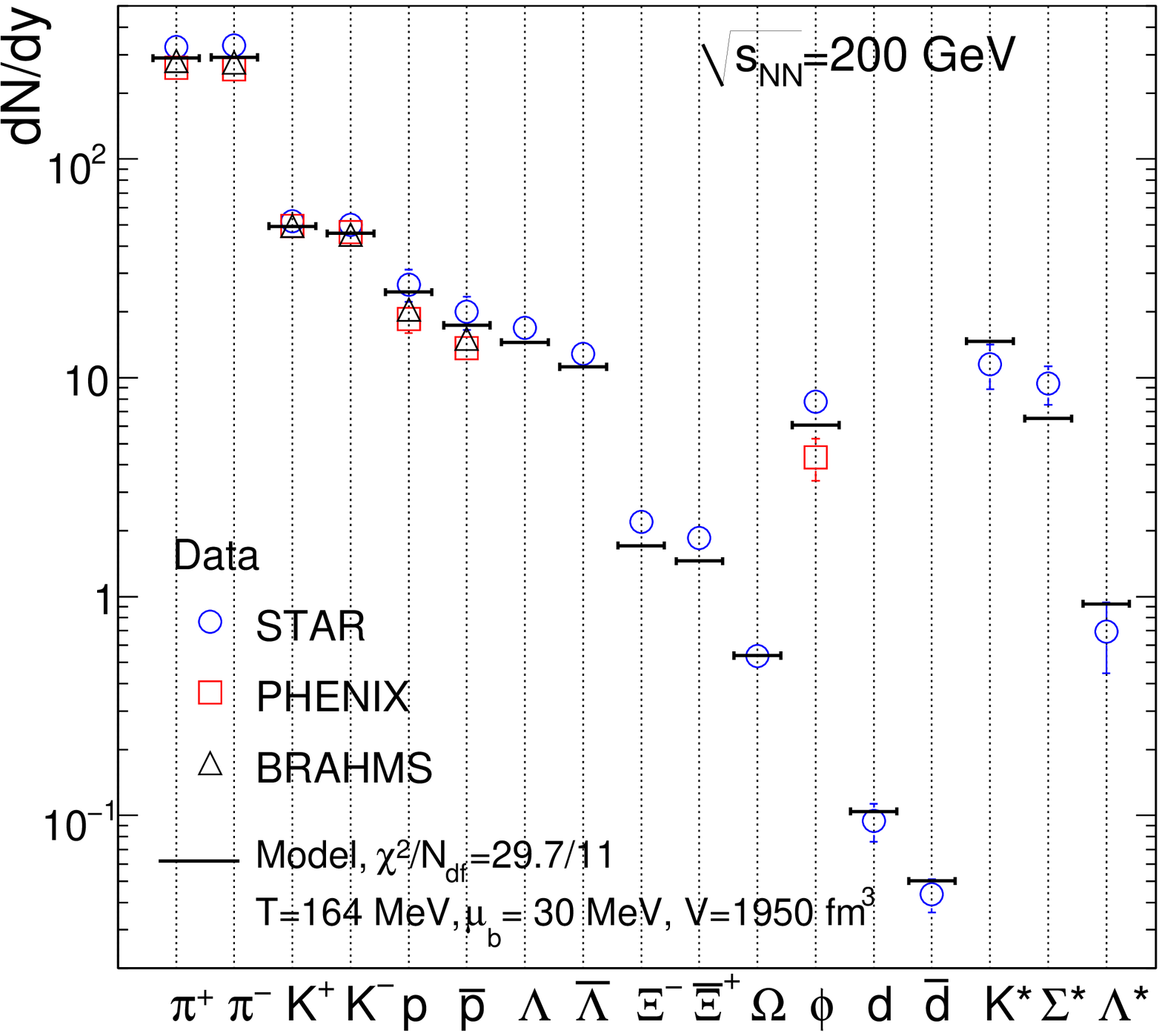}
\end{minipage} 
\end{tabular}
\caption{Experimental hadron yields and model calculations for the parameters 
of the best fit at the energies of 7.6 (left panel) and 200 GeV (right panel;
the $\Omega$ yield includes both $\Omega^-$ and $\bar{\Omega}^+$).}
\label{fig_fits}
\vspace{-.5cm}
\end{figure}

In Fig.~\ref{fig_fits} we present a comparison of measured and calculated
hadron yields at the energies of  
$\sqrt{s_{NN}}$=7.6 GeV (beam energy of 30 AGeV at SPS) and 
$\sqrt{s_{NN}}$=200 GeV. The model is successful in reproducing 
the measurements and this applies to all energies, from 2 AGeV beam energy
(fixed target) up to the top RHIC energy of $\sqrt{s_{NN}}$=200 GeV. 
The reduced $\chi^2$ values are reasonable.
In most cases the fit quality is improved compared to our earlier analysis 
\cite{aa05}, even though the experimental errors are now smaller.
A disagreement between the experiments is seen at the top RHIC energy for 
pions and protons, see Fig.~\ref{fig_fits}, which is the reason of the 
large reduced $\chi^2$. A fit of ratios is in this case more suited,
but we note that a fit of the STAR yields alone gives $T$=162 MeV, $\mu_b$=32 MeV, 
$V$=2400 fm$^3$, with a very good $\chi^2/N_{df}$=9.0/11. The resonances were 
not included in the fits, but are quite well reproduced by the model.

\begin{figure}[hbt]
\centering\includegraphics[width=1.\textwidth]{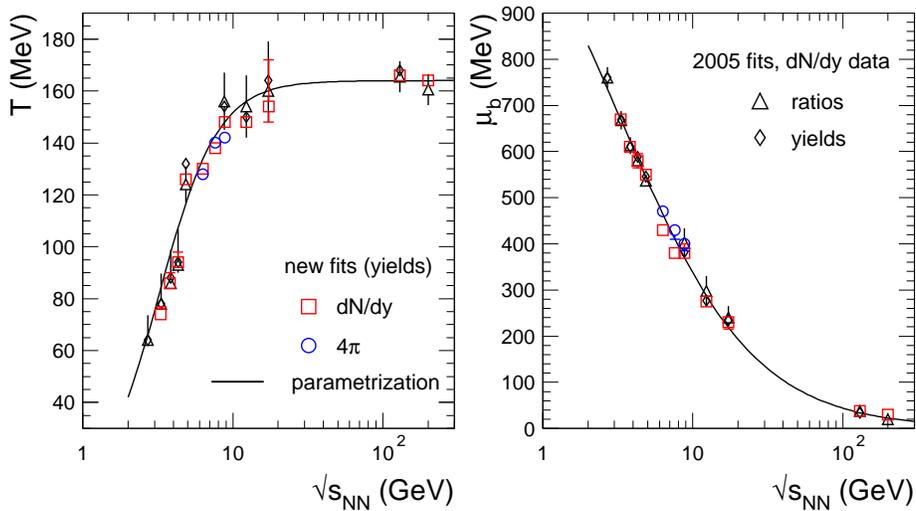}
\caption{The energy dependence of the temperature and baryon chemical potential 
at chemical freeze-out. The  results of the new fits \cite{aat2} are compared 
to the values obtained in our earlier study \cite{aa05}.
The lines are parametrizations for $T$ and $\mu_b$ (see text).}
\label{fig_tmu}
\end{figure}

An important result of our analysis is that the resulting thermal parameters
are close to those obtained earlier \cite{aa05} and are in agreement 
with other recent studies \cite{man08,star08}.  
This indeed confirms that the common practice of including in the thermal codes 
hadrons up to masses of 2 GeV (for instance in the publicly-available code 
THERMUS \cite{thermus}) does not lead to significantly biased fit parameters.
Nevertheless, there are small variations.
In Fig.~\ref{fig_tmu} we present the energy dependence of $T$ and $\mu_b$ 
in comparison to our earlier results \cite{aa05}.
We have parametrized $T$ as a function 
of $\sqrt{s_{NN}}$ with the following expression\footnote{For $\mu_b$, 
  there is no need to change our earlier\cite{aa05} parametrization:  
$\mu_b \mathrm{[MeV]}=\frac{1303}{1+0.286\sqrt{s_{NN}(\mathrm{GeV})}}$}:
\be
T \mathrm{}=T_{lim}\frac{1}{1+\exp(2.60-\ln(\sqrt{s_{NN}(\mathrm{GeV})})/0.45)},
\label{pt}
\ee
with the "limiting" temperature $T_{lim}$=164 MeV. This value is slightly 
higher compared to our earlier value of 161$\pm$4 MeV \cite{aa05} due to 
the higher temperatures presently derived for the RHIC energies.
The approach to $T_{lim}$ is presently more gradual compared to our earlier
parametrization.

The values of $\mu_b$ extracted for the two lowest SPS energies deviate
somewhat from the continuous trend suggested by all the other points.  At these
energies the fit to data in full phase space does lead, as expected, to larger
values of $\mu_b$, which do fit in the systematics. At 40 AGeV
($\sqrt{s_{NN}}$=8.8 GeV) the resulting values of $T$ and $\mu_b$ from the fit
of 4$\pi$ data are very similar to those obtained from midrapidity data.

\begin{figure}[hbt]
\centering\includegraphics[width=.65\textwidth]{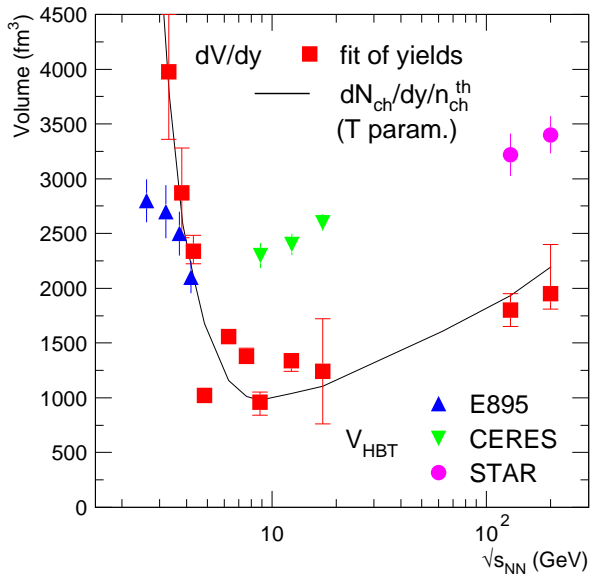}
\caption{The fireball volume at chemical freeze-out compared to the volume 
extracted from the HBT measurements \cite{cer}.}
\label{fig_vol}
\end{figure}

The volume at chemical freeze-out (corresponding to a slice of one unit 
of rapidity, $\mathrm{d}V/\mathrm{d}y$) is shown in Fig.~\ref{fig_vol} as 
a function of energy.
The values extracted directly from the fits of particle yields 
are compared to the values obtained by dividing the measured charged particle 
rapidity densities with densities calculated employing the above parametrizations
of $T$ and $\mu_b$. As expected, the two methods give very similar results.
The chemical freeze-out volume is compared to the kinetic freeze-out volume 
extracted from Hanbury Brown and Twiss (HBT) measurements, $V_{HBT}$ \cite{cer}.
Note that, to relate quantitatively the magnitude of the rapidity density 
of the chemical freeze-out volume $\mathrm{d}V/\mathrm{d}y$ to the volume 
at kinetic freeze-out determined from HBT measurements, one needs to map rapidity 
onto space (see the discussion in \cite{cer}). 
Here we only observe that the energy dependence of the two observables
exhibit a similar non-monotonic behavior.
It is interesting to note that the minimum of $\mathrm{d}V/\mathrm{d}y$ 
occurs at the $\sqrt{s_{NN}}$ corresponding to the saturation of $T$.

\begin{figure}[htb]
\begin{tabular}{lr} \begin{minipage}{.49\textwidth}
\hspace{-.3cm}\includegraphics[width=1.\textwidth]{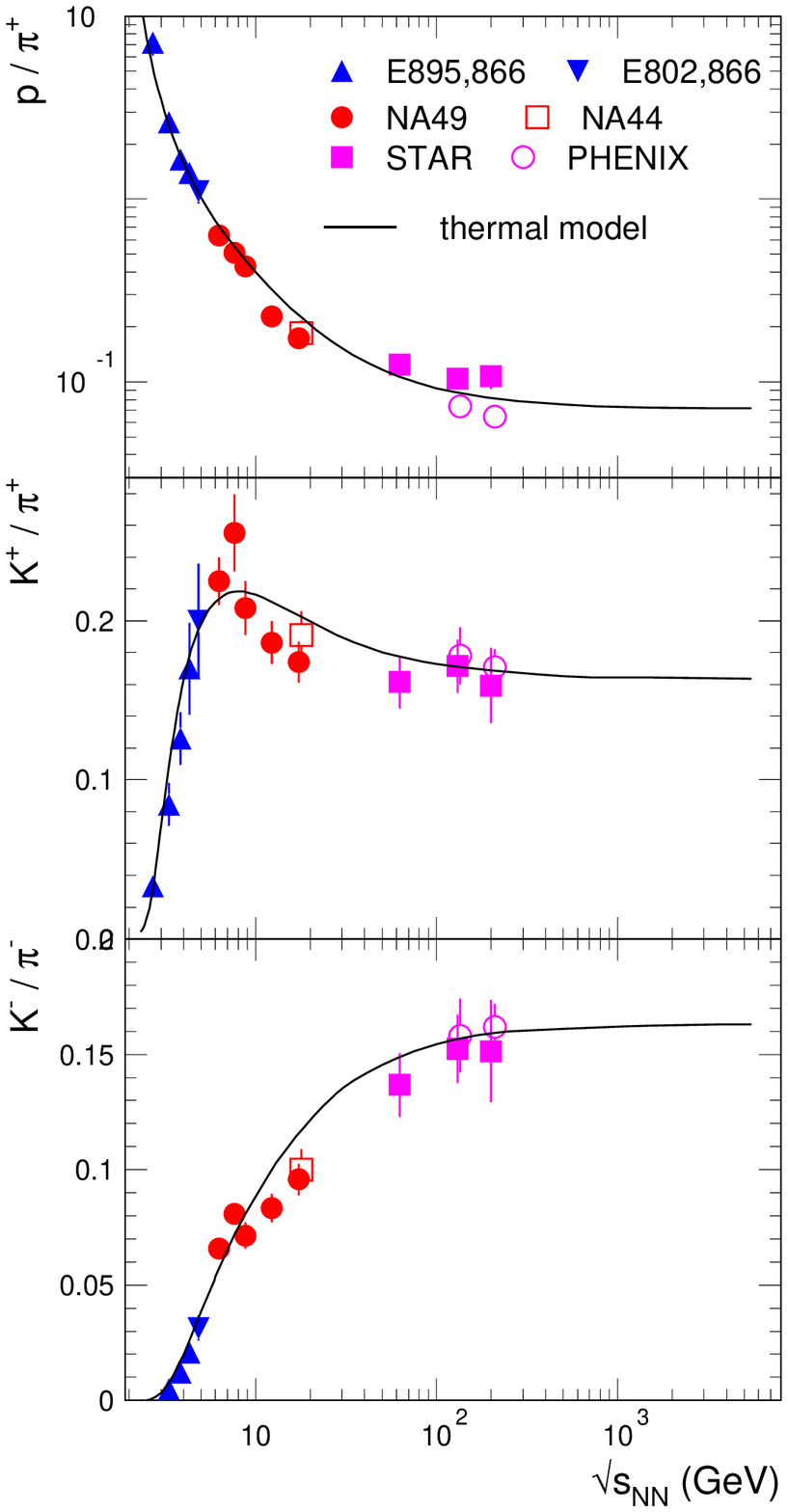}
\end{minipage} &\begin{minipage}{.49\textwidth}
\hspace{-.5cm}\includegraphics[width=1.\textwidth]{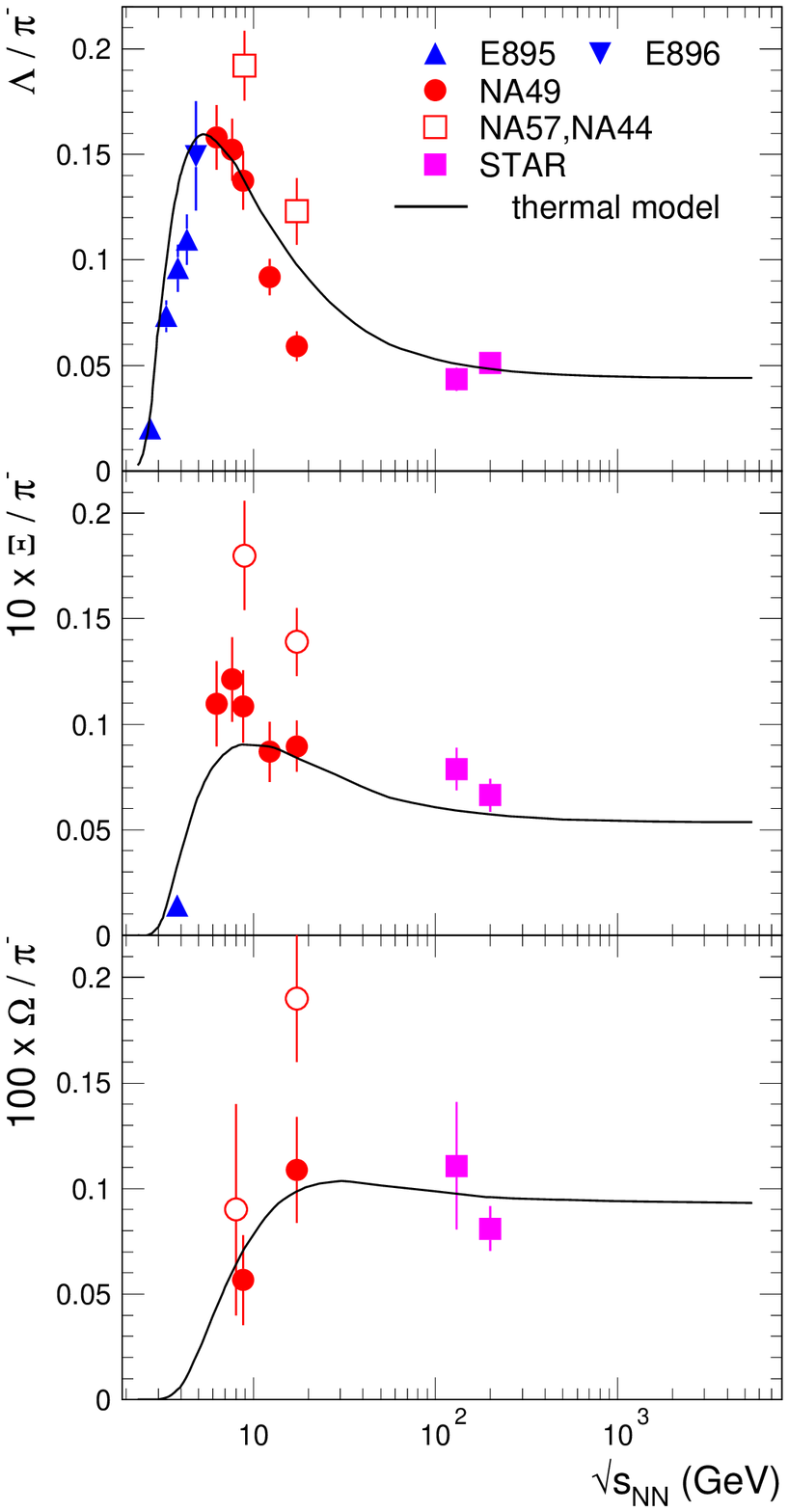}
\end{minipage} \end{tabular}
\caption{Energy dependence of hadron yields relative to pions.}
\label{fig_k2pi}
\end{figure}

We employ the above parametrizations of $T$ and $\mu_b$ to investigate the 
energy dependence of the production yields of various hadrons relative to pions,
shown in Fig.~\ref{fig_k2pi}.
In particular, the $K^+/\pi^+$ ratio shows a rather pronounced maximum 
at a beam energy of 30 AGeV \cite{na49pi}, and the data are well reproduced by 
the model calculations.
In the thermal model this maximum occurs naturally at 
$\sqrt{s_{NN}}\simeq$8 GeV \cite{pbm4}. It is due to the counteracting effects
of the steep rise and saturation of $T$ and the strong monotonous decrease 
in $\mu_b$.
The competing effects are most prominently reflected in the energy dependence
of the $\Lambda$ hyperon to pion ratio (right panel of Fig.~\ref{fig_k2pi}), 
which shows a pronounced maximum at $\sqrt{s_{NN}}\simeq$5 GeV. 
This is reflected in the $K^+/\pi^+$ ratio somewhat less directly; it appears 
mainly as a consequence of strangeness neutrality, assumed in our calculations.

The model describes the $K^+/\pi^+$ data very well over the full energy range,
as a consequence of the inclusion in the code of the high-mass resonances and 
of the $\sigma$ meson, while our earlier calculations \cite{aa05} were 
overpredicting the SPS data. At RHIC energies, the quality of the present fits 
is essentially unchanged compared to \cite{aa05}, as also the data have changed
somewhat.
The model also describes accurately the $\Lambda/\pi^-$ measurements
as well as those for other hyperons.
We note that the maxima in the various production ratios are located at 
different energies \cite{cleymans04}. The model calculations reproduce this 
feature in detail.

The calculated $K^+/\pi^+$ ratio is likely to decrease further at energies beyond 
the maximum and the peak is likely to sharpen somewhat \cite{aat2} if our presumably 
still incomplete knowledge of the hadronic spectrum for masses larger than 2 GeV 
would improve.
The uncertainty of the calculations due to the mass and width of the $\sigma$ 
meson are at the percent level only. 
Another few percent uncertainty, which is difficult to
assess quantitatively, arises from the unknown branching ratios of the 
high-mass resonances.

In summary, our results \cite{aat2} demonstrate that by inclusion of 
the $\sigma$ meson and many higher mass resonances into the resonance spectrum 
employed in the statistical model calculations an improved description is 
obtained of hadron production in central nucleus-nucleus collisions at 
ultra-relativistic energies. A dramatic improvement is visible 
for the $K^+/\pi^+$ ratio, which is now well described at all energies. 
The ``horn'' finds herewith a natural explanation which is, however, 
deeply rooted in and connected to detailed features of the hadronic mass 
spectrum which leads to a limiting temperature and contains the QCD phase 
transition \cite{hagedorn85}. 
Our results strongly imply that hadronic observables near and above the horn 
structure at a beam energy of 30 AGeV provide a link
to the QCD phase transition. Open questions are whether the chemical
freeze-out curve below the horn energy actually traces the QCD phase boundary
at large values of chemical potential or whether chemical freeze-out in this
energy range is influenced by exotic new phases such as have been predicted in
\cite{mclerran_pisarski}. In any case these are exciting prospects for new
physics to be explored at RHIC low energy runs and, in particular, at the high
luminosity FAIR facility. 
At the high energy frontier, the measurements at LHC will be a crucial test 
of the present picture.

\vspace{.2cm}
Acknowledgements:   
We acknowledge the support from the Alliance Program of the Helmholtz-Gemeinschaft.

\end{document}